\documentclass[conference]{IEEEtran}
\IEEEoverridecommandlockouts
\usepackage{cite}
\usepackage{amsmath,amssymb,amsfonts}
\usepackage{algorithmic}
\usepackage{graphicx}
\usepackage{textcomp}
\usepackage{xcolor}
\def\BibTeX{{\rm B\kern-.05em{\sc i\kern-.025em b}\kern-.08em
    T\kern-.1667em\lower.7ex\hbox{E}\kern-.125emX}}
\usepackage{tikz}
\usepackage[capitalise,noabbrev]{cleveref}
\usepackage{booktabs}
\usepackage{siunitx}
\DeclareSIQualifier\isotropic{i}
\newcommand{\comment}[1]{}

\usepackage[printonlyused,nolist]{acronym}

\AtBeginDocument{

\begin{acronym}

    \acro{4G}[4G]{4th generation}
    \acro{5G}[5G]{5th generation}
    \acro{6G}[6G]{6th generation}
    \acro{5GPPP}[5G-PPP]{5G infrastructure public private partnership}
    \acro{3GPP}[3GPP]{3rd Generation Partnership Project}
    \acro{ADC}[ADC]{analog-to-digital converter}
    \acro{APSM}[APSM]{adaptive projected subgradient method}
    \acro{APP}[APP]{a posteriori probability}
    \acro{ASIC}[ASIC]{Application-Specific Integrated Circuit}
    \acro{AWGN}[AWGN]{additive white Gaussian noise}
    \acro{BER}[BER]{bit error ratio}
    \acro{BMI}[BMI]{bitwise mutual information}
    \acro{BPSK}[BPSK]{binary phase-shift keying}
    \acro{BS}[BS]{base station}
    \acro{CBER}[CBER]{coded bit error ratio}
    \acro{CFO}[CFO]{carrier frequency offset}
    \acro{COMP}[CoMP]{coordinated multi-point}
    \acro{COTS}[COTS]{commercial off-the-shelf}
    \acro{CPU}[CPU]{central processing unit}
    \acro{DAC}[DAC]{digital-to-analog converter}
    \acro{DL}[DL]{downlink}
    \acro{DMA}[DMA]{direct memory access}
    \acro{DU}[DU]{Distributed Unit}
    \acro{EC}[EC]{European commission}
    \acro{eNB}[eNB]{evolved Node B}
    \acro{FPGA}[FPGA]{field-programmable gate array}
    \acro{FR1}[FR1]{Frequency Range 1}
    \acro{FR2}[FR2]{Frequency Range 2}
    \acro{FEC}[FEC]{forward error correction} 
    \acro{GEMM}{general matrix multiplication}
    \acro{gNB}[gNB]{next generation Node B}
    \acro{GPS}[GPS]{global positioning system}
    \acro{GPU}[GPU]{graphics processing unit}
    \acro{H2020}[H2020]{Horizon 2020}
    \acro{HPC}[HPC]{high-performance computing}
    \acro{IIO}[IIO]{industrial I/O}
    \acro{IOT}[IoT]{internet of things}
    \acro{LDPC}{low-density parity-check}
    \acro{LLR}{log likelihood ratio}
    \acro{LLS}[LLS]{linear least squares}
    \acro{LO}[LO]{local oscillator}
    \acro{LTE}[LTE]{long-term evolution}
    \acro{MAP}[MAP]{maximum a-posteriori}
    \acro{MCM}[MCM]{multi-carrier modulation}
    \acro{ML}[ML]{machine learning}
    \acro{MIMO}[MIMO]{multiple-input multiple-output}
    \acro{MMIMO}[mMIMO]{massive multiple-input multiple-output}
    \acro{MMSE}[MMSE]{minimum mean square error}
    \acro{MMTC}[mMTC]{massive machine-type communication}
    \acro{MSE}{mean squared error}
    \acro{NOMA}[NOMA]{non-orthogonal multiple access}
    \acro{MU}[MU]{multiuser}
    \acro{MA}[MA]{multiple access}
    \acro{MUSA}[MUSA]{multi-user Shared Access}
    \acro{NI}[NI]{National Instruments}
    \acro{NN}{neural network}
    \acro{NL}[NL]{nonlinear}
    \acro{NR}[NR]{New Radio}
    \acro{NV}[NV]{Nvidia}
    \acro{OFDM}[OFDM]{orthogonal frequency-division multiplexing}
    \acro{OFDMA}[OFDMA]{orthogonal frequency-division multiple access}
    \acro{OMA}[OMA]{orthogonal multiple access}
    \acro{OTB-5G+}[OTB-5G+]{Open Testbed Berlin - 5G and Beyond}
    \acro{PC}[PC]{personal computer}
    \acro{PLL}[PLL]{phase lock loop}
    \acro{QAM}{quadrature amplitude modulation}
    \acro{QPSK}[QPSK]{quadrature phase-shift keying}
    \acro{RAN}[RAN]{Radio Access Network}
    \acro{ReLU}[ReLU]{rectified linear unit}
    \acro{RKHS}[RKHS]{reproducing kernel Hilbert space}
    \acro{RRH}[RRH]{remote radio head}
    \acro{Rx}[Rx]{receiver}
    \acro{RRC}[RRC]{root-raised-cosine}
    \acro{SCM}[SCM]{Single-Carrier Modulation}
    \acro{SC-FDMA}[SC-FDMA]{Single-Carrier Frequency-Division Multiple Access}
    \acro{SDR}[SDR]{Software-Defined Radio}
    \acro{SGD}[SGD]{stochastic gradient descent}
    \acro{SIC}[SIC]{successive interference cancellation}
    \acro{SIMO}[SIMO]{Single-Input Multiple-Output}
    \acro{SIMT}[SIMT]{single instruction multiple threads}
    \acro{SINR}[SINR]{signal-to-interference-plus-noise ratio}
    \acro{SNR}[SNR]{signal-to-noise ratio}
    \acro{SoC}[SoC]{System-on-Chip}
    \acro{SOI}[SOI]{signal of interest}
    \acro{SER}[SER]{symbol error ratio}
    \acro{TS}[TS]{technical specification}
    \acro{Tx}[Tx]{transmitter}
    \acro{UE}[UE]{user equipment}
    \acro{UCA}[UCA]{uniform circular array}
    \acro{UL}[UL]{uplink}
    \acro{ULL}[ULL]{ultra-low latency}
    \acro{URLLC}[URLLC]{ultra-reliable and low latency communications}
    \acro{USRP}[USRP]{Universal Software Radio Peripheral}
    \acro{UHD}[UHD]{USRP Hardware Driver}
    \acro{UPA}[UPA]{Uniform Planar Array}
    \acro{WLAN}[WLAN]{Wireless Local Area Network}
    \acro{CUDA}[CUDA]{Compute Unified Device Architecture}
    \acro{GPC}[GPC]{Graphics Processing Cluster}
    \acro{TPC}[TPC]{Texture Processing Cluster}
    \acro{SM}[SM]{Streaming Multiprocessor}
    \acro{RT}[RT]{ray tracing}
    \acro{TDP}[TDP]{Thermal Design Power}
    \acro{GDDR}[GDDR]{Graphics Double Data Rate}
    \acro{HBM}[HBM]{High Bandwidth Memory}
    \acro{MATLAB}[MATLAB]{MATrix LABoratory}
    
\end{acronym}

}

\usepackage{xspace}

\makeatletter
\DeclareRobustCommand\onedot{\futurelet\@let@token\@onedot}
\def\@onedot{\ifx\@let@token.\else.\null\fi\xspace}

\makeatother

\newcommand{\iu}{\mathrm{i}\mkern1mu}

\renewcommand{\baselinestretch}{0.995}

\begin{document}

\title{GPU-Accelerated Machine Learning in Non-Orthogonal Multiple Access\\
\thanks{This work has been partially funded by the German Federal Ministry of Education and Research (BMBF, Germany) in the project Open Testbed Berlin - 5G and Beyond (OTB-5G+) under Grant 16KIS0980.}
}

\author{\IEEEauthorblockN{Daniel Schäufele}
\IEEEauthorblockA{
\textit{Fraunhofer Heinrich Hertz Institute}\\
Berlin, Germany \\
daniel.schaeufele@hhi.fraunhofer.de}
\and
\IEEEauthorblockN{Guillermo Marcus}
\IEEEauthorblockA{
\textit{NVIDIA}\\
Berlin, Germany \\
gmarcus@nvidia.com}
\and
\IEEEauthorblockN{Nikolaus Binder}
\IEEEauthorblockA{
\textit{NVIDIA}\\
Berlin, Germany \\
nbinder@nvidia.com}
\and[\hfill\mbox{}\par\hfill]
\IEEEauthorblockN{Matthias Mehlhose}
\IEEEauthorblockA{
\textit{Fraunhofer Heinrich Hertz Institute}\\
Berlin, Germany \\
matthias.mehlhose@hhi.fraunhofer.de}
\and
\IEEEauthorblockN{Alexander Keller}
\IEEEauthorblockA{
\textit{NVIDIA}\\
Berlin, Germany \\
akeller@nvidia.com}
\and
\IEEEauthorblockN{S{\l}awomir Sta{\'n}czak}
\IEEEauthorblockA{
\textit{Fraunhofer Heinrich Hertz Institute}\\
Berlin, Germany \\
slawomir.stanczak@hhi.fraunhofer.de}
}

\maketitle

\begin{abstract}
Non-orthogonal multiple access (NOMA) is an interesting technology that enables massive connectivity as required in future 5G and 6G networks. While purely linear processing already achieves good performance in NOMA systems, in certain scenarios, non-linear processing is mandatory to ensure acceptable performance.

In this paper, we propose a neural network architecture that combines the advantages of both linear and non-linear processing. Its real-time detection performance is demonstrated by a highly efficient implementation on a graphics processing unit (GPU).  Using real measurements in a laboratory environment, we show the superiority of our approach over conventional methods.
\end{abstract}


\section{Introduction}

In future 5G and 6G mobile networks, the demand for massive connectivity will not be satisfiable with traditional \ac{OMA} systems. This is especially the case in \ac{MMTC} scenarios, e.g., in campus networks. For this reason a large body of research has been devoted to \ac{NOMA} systems\cite{Shin2017,Wang2016,Ding2017,Tabassum2016}. In order to deal with multi-user interference in such systems, non-linear detection has been proposed (for example, see \cite{Xin2016,Islam2017,Awan_thesis}).

While non-linear detectors can perform significantly better than linear detectors in scenarios with strong multi-user interference, they can be very sensitive to even small changes in a wireless environment (e.g., due to multi-path scattering and intermittent interference in \ac{MMTC} scenarios).
Therefore, the performance of non-linear detectors can degrade in dynamic environments.
Theoretical studies \cite{Bjornson2017} have shown that in massive \ac{MIMO} systems linear detectors can achieve the spectral efficiency comparable to non-linear methods. For this reason, purely non-linear detectors may be inefficient in massive \ac{MIMO} \ac{NOMA} systems.

Contrary to our previous work which focused on using the \ac{APSM}\cite{mehlhose2022real}, we propose a \ac{NN} architecture, which combines a linear and a non-linear branch. Additionally, we propose to use the \ac{LLS} algorithm to initialize the weights of the linear branch and to exploit symmetry in the IQ samples to improve the performance over a conventional \ac{NN}. We then present a \ac{GPU}-based implementation, which is able to achieve very short execution times. In addition, we evaluate the performance on a real data set that has been acquired in a lab environment.

The remainder of this paper is organized as follows. In \cref{sec:background} we give a formal problem statement as well as a presentation of our proposed \ac{NN} architecture and several tricks we used to increase the performance. In \cref{sec:implementation,sec:experimental_setup} we present the real-time implementation and the laboratory setup, which led to the results discussed in \cref{sec:results}.

\section{Background} \label{sec:background}
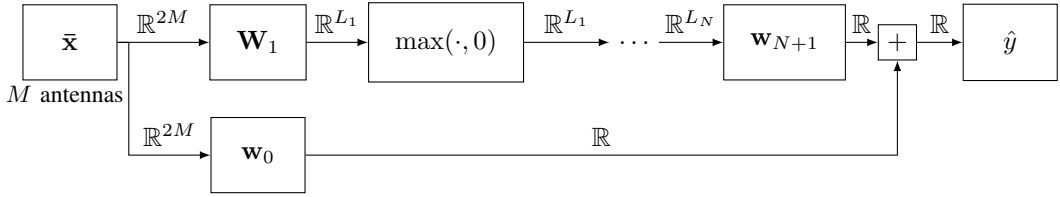
\begin{figure*}
    \centering
    \begin{tikzpicture}[box/.style={draw, rectangle, minimum width = 1.25cm, minimum height = 1cm, inner sep = 10}] 
       \node[box] (input) at (0, 0) { $\mathbf{\bar{x}}$ };
       \node[font = \small, yshift = -5, xshift = -2] at (input.south) {$M$ antennas};
       \node[box] (l1) at (2.5, 0) { $\mathbf{W}_1$ };
       \node[box] (l2) at (5, 0) { $\max(\cdot, 0)$ };
       \node (ldots) at (7.5, 0) { $\ldots$ };
       \node[box] (l3) at (9.5, 0) { $\mathbf{w}_{N+1}$ };
       \node[box] (j1) at (2.5, -1.5) { $\mathbf{w}_0$ };
       \node[draw] (add) at (11, 0) { $ + $};
       \node[box] (output) at (12.5, 0) { $\hat{y}$ };
       \draw[-latex] (input.east) -- (l1.west) node [pos = 0.5, above] { $\mathbb{R}^{2M}$ };
       \draw[-latex] (input.east) -- ++(0.15, 0) -- ++(0, -1.5) -- (j1.west) node [pos = 0.5, above] { $\mathbb{R}^{2M}$ };
       \draw[-latex] (l1.east) -- (l2.west) node [pos = 0.5, above] { $\mathbb{R}^{L_1}$ };
       \draw[-latex] (l2.east) -- (ldots.west) node [pos = 0.5, above] { $\mathbb{R}^{L_1}$ };
       \draw[-latex] (ldots.east) -- (l3.west) node [pos = 0.5, above] { $\mathbb{R}^{L_N}$ };
       \draw[-latex] (l3.east) -- (add.west) node [pos = 0.5, above] { $\mathbb{R}$ };
       \draw[-latex] (add.east) -- (output.west) node [pos = 0.5, above] { $\mathbb{R}$ };
       \draw[latex-] (add.south) -- (add |- 0, 0 |- j1) -- (j1.east) node [pos = 0.5, above] { $\mathbb{R}$ };
    \end{tikzpicture}
    \vspace{-3mm}
    \caption{Proposed structure of \ac{NN} with $N$ hidden layers.}
    \label{fig:nn_structure}
\end{figure*}
In this section we first give an overview of the system model and define the problem we intend to solve. Then we present our \ac{NN} architecture and several tricks to improve its performance.

\subsection{Preliminaries}

In the following, scalars, column vectors and matrices are denoted by italic lower case letters $x$, bold lower case letters $\mathbf{x}$ and bold upper case letters $\mathbf{X}$, respectively. Matrix transposes and inverses are denoted by $\mathbf{X}^\mathsf{T}$ and $\mathbf{X}^{-1}$, respectively. The set of natural numbers, real numbers and complex numbers are denoted by $\mathbb{N}$, $\mathbb{R}$ and $\mathbb{C}$, respectively, while the real and imaginary parts of a complex number $c \in \mathbb{C}$ are denoted by $\Re(c)$ and $\Im(c)$, respectively.
We define the range $\overline{N_{1},N_{2}}:=\left\{N_1,N_{1}+1,\ldots,N_2\right\} \subset \mathbb{N}$, where $N_1\leq N_2$. 

\subsection{System Model}

We assume a multi-user uplink system with $K$ users and $M$ receive antennas and a non-dispersive channel. The received signal $\mathbf{r}(t) \in \mathbb{C}^M$ at the time $t \in \mathbb{N}$ is given by 
\begin{equation}
	\mathbf{r}(t) = \sum_{k=1}^{K}\sqrt{p_{k}} \cdot b_{k}(t) \cdot \mathbf{h}_{k}+ \mathbf{n}(t),
	\label{eqn:uplink_signal}
\end{equation} 
where $p_{k}\in\mathbb{R}$ is the transmit power of user $k \in \overline{1,K}$, $b_{k}(t)\in\mathbb{C}$ is its information-bearing symbol, while the vectors $\mathbf{h}_{k}\in\mathbb{C}^M$ and $\mathbf{n}(t)\in\mathbb{C}^M$ stand for the channel signature of user $k$ and additive noise, respectively.
Note that we do not assume any distribution of the noise and structure of the receive antenna array.
The objective of multi-user detection considered in this study is to design a  filter $g^{k}:\mathbb{C}^{M} \to \mathbb{C}$ for a selected user $k$, such that $(\forall t\in\mathbb{N})$ $\left|g^{k}(\mathbf{r}(t))-{b}_{k}(t)\right|\leq\epsilon$, where $\epsilon>0$ is a small predefined noise tolerance.

The receive process is split into two phases. In the first (training) phase all transmitters transmit a series of $N_\mathrm{T}$ uniformly distributed pseudo-random symbols, which are known by the receiver. The transmitted symbols for user $k$ are then collected into a vector $\mathbf{y}_{\mathrm{T},k} = \left[b_{\mathrm{T},k}(1), b_{\mathrm{T},k}(2), \ldots, b_{\mathrm{T},k}(N_\mathrm{T})\right]^\mathsf{T} \in \mathbb{C}^{N_\mathrm{T}}$. The corresponding receive signal is also collected into a matrix $\mathbf{X}_\mathrm{T} = \left[\mathbf{r}(1), \mathbf{r}(2), \ldots, \mathbf{r}(N_\mathrm{T}) \right]^\mathsf{T} \in \mathbb{C}^{N_\mathrm{T} \times M}$. This training data can then be utilized to train the detection filter.

During the following detection phase a series of $N_\mathrm{D}$ data symbols is transmitted by each transmitter, which is to be reconstructed at the receiver. Again, we collect the receive signal into a matrix $\mathbf{X}_\mathrm{D} = \left[\mathbf{r}(N_\mathrm{T} + 1), \mathbf{r}(N_\mathrm{T} + 2), \ldots, \mathbf{r}(N_\mathrm{T} + N_\mathrm{D}) \right]^\mathsf{T} \in \mathbb{C}^{N_\mathrm{D} \times M}$ and use this matrix as input for the detection algorithm.

\subsection{Exploiting the Symmetry of IQ Samples}\label{sec:complex_symmetry}

According to the argument given in \cite{slavakis2009adaptive}, we can use the equivalence
\begin{equation}
    g(\mathbf{r}(t)) = f(\mathbf{r_1}(t)) + \iu f(\mathbf{r_2}(t)) \label{eq:complex_symmetry}
\end{equation}
for linear functions $f : \mathbb{R}^{2M} \to \mathbb{R}$ and $g: \mathbb{C}^M \to \mathbb{C}$, where $\mathbf{r_1}(t) = \left[\Re(\mathbf{r}(t))^\mathsf{T}, \Im(\mathbf{r}(t))^\mathsf{T} \right]^\mathsf{T} \in \mathbb{R}^{2M}$ and $\mathbf{r_2}(t) = \left[\Im(\mathbf{r}(t))^\mathsf{T}, -\Re(\mathbf{r}(t))^\mathsf{T} \right]^\mathsf{T} \in \mathbb{R}^{2M}$. Applying this transformation to the training input matrix $\mathbf{X}_\mathrm{T}$ yields a new matrix $\mathbf{\bar{X}}_\mathrm{T} = \left[\mathbf{r_1}(1), \mathbf{r_2}(1), \ldots, \mathbf{r_1}(N_\mathrm{T}), \mathbf{r_2}(N_\mathrm{T}) \right]^\mathsf{T} \in \mathbb{R}^{2N_\mathrm{T} \times 2M}$ and a new training target vector $\mathbf{\bar{y}}_{\mathrm{T},k} = \left[\Re(b_k(1)), \Im(b_k(1)), \ldots, \Re(b_k(N_\mathrm{T})), \Im(b_k(N_\mathrm{T})) \right]^\mathsf{T} \in \mathbb{R}^{2N_\mathrm{T}}$. The same transformation can be applied to the detection input matrix $\mathbf{X}_\mathrm{D}$ to obtain $\mathbf{\bar{X}}_\mathrm{D} \in \mathbb{R}^{2N_\mathrm{D} \times 2M}$ and $\mathbf{\bar{y}}_{\mathrm{D},k} \in \mathbb{R}^{2N_\mathrm{D}}$.

By using this approach, we can use twice as many training samples to train $f$. At the same time we enforce that both real and imaginary parts are predicted using the same function (with different inputs), which decreases the degrees of freedom of the learned function (compared to using two independent functions). This will speed up the training process given that the constraint is satisfied by the input data.

We later show experimentally that this approach is also beneficial in the case of non-linear \acp{NN}, although there is no strong theoretical background for this, yet.

\subsection{Neural Network Structure}\label{sec:nn_structure}

In many cases a purely linear estimator can already achieve good performance. However, in certain scenarios, non-linear processing (which we will provide with \acp{NN}) is necessary to achieve good performance. In order to combine both approaches, we propose to use the architecture shown in \cref{fig:nn_structure}. The linear (bottom) branch provides a dense layer without any activation functions, while the top branch contains $N$ hidden, dense layers with $L_n$ neurons each and \ac{ReLU} activation functions, where $n \in \overline{1, N}$, followed by a final dense layer without non-linearity. Both branches are summed element-wise to produce the final output. This approach resembles a single residual unit in deep residual networks (ResNets)\cite{he2016identity}. However, in the proposed architecture, the skip connection contains a matrix multiplication, whose weights we efficiently initialize in a manner outlined in the next section. Due to this initialization, explicit orthogonalization of both branches is unnecessary, as the non-linear branch refines the output of the linear estimator.

\subsection{\Ac{LLS} Initialization of Linear Weights}\label{sec:lls_initialization}

The optimal solution (under certain assumptions) for a linear estimator can be computed using the \ac{LLS} algorithm\cite[p. 83]{luenberger1997optimization}, where the weights for user $k$ are computed as
\begin{equation}
    \mathbf{w}^k = (\mathbf{\bar{X}}_\mathrm{T}^\mathsf{T}\mathbf{\bar{X}}_\mathrm{T})^{-1} \mathbf{\bar{X}}_\mathrm{T}^\mathsf{T} \mathbf{\bar{y}}_{\mathrm{T},k}.
\end{equation}

We can then compute the solution by a matrix-vector multiplication as $\hat{\mathbf{y}}_{\mathrm{D},k} =  \mathbf{\bar{X}}_\mathrm{D} \mathbf{w}^k$. By using the weights $\mathbf{w}^k$ for the weight vector $\mathbf{w}_0$ in the linear branch of the \ac{NN} presented in \cref{sec:nn_structure}, we can incorporate the \ac{LLS} solution into our \ac{NN}. This approach is beneficial to reduce training time as the training of the \ac{NN} by traditional \ac{SGD} methods already has a good starting point. The number of computations needed for training is further reduced by fixing the weights of the linear branch and thus avoiding the need for weight updates, resulting in a key performance improvement.

Under certain conditions, we can achieve better performance with \ac{MMSE} estimation\cite{verdu1998multiuser}, but we deem the added computational effort to be unnecessary, as the estimator will be further refined during the training of the neural network.

\section{Real-Time Implementation} \label{sec:implementation}

In order to achieve real-time performance for our proposed \ac{NN} structure we propose to use an optimized implementation running on a \ac{GPU}. For the purpose of this work we consider the fully fused implementation that was used in \cite{muller2021real}. This implementation is heavily optimized to save all weights of the networks into the extremely fast registers and to save all intermediate results into shared memory, thus avoiding the need to access the relatively slow global memory. Additionally, the full network is fused into a single kernel, which avoids the need to copy intermediate results to global memory. However, due to the limited amount of registers and shared memory, this approach only works for up to 128 neurons per layer. To show results for layers with a larger number of neurons, we switch to an implementation that applies the \ac{GEMM} routines provided by the CUTLASS template library when necessary.

In order to optimize the computation of the \ac{LLS} solution we used the \texttt{cublasSgelsBatched} algorithm provided by the cuBLAS library.

\section{Experimental Setup} \label{sec:experimental_setup}
This section describes the hardware components and transmission signals of our real-time multi-user detection setup. 

\subsection{Hardware Components}

\begin{figure}[t] 
    \centering
    \includegraphics[width=\linewidth]{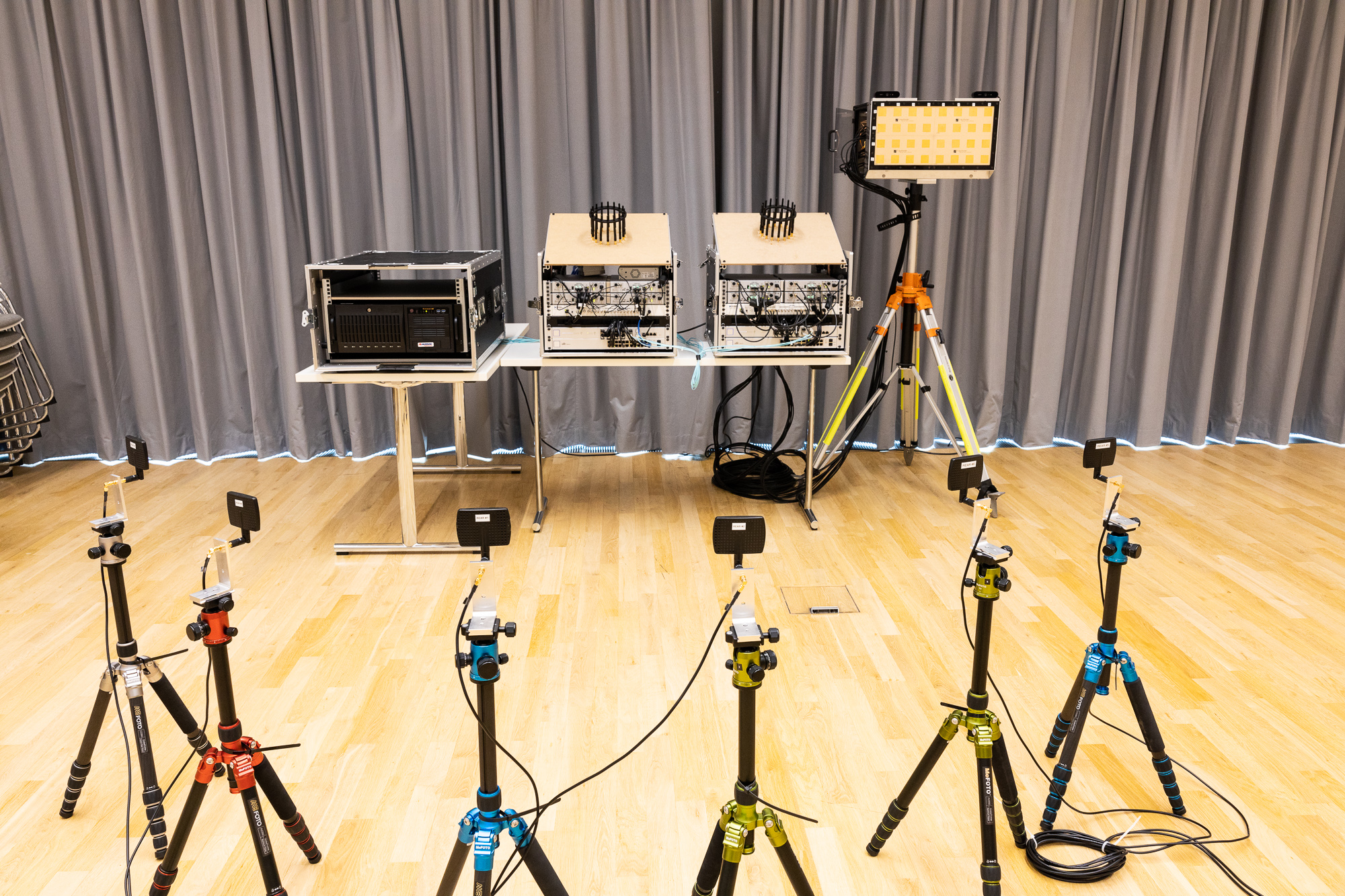}
    \vspace{-7mm}
    \caption{System setup. The antennas in the front are used as transmitters, the receiver consists of one the circular antenna array in the background (only one is used for theses experiments), the Ettus \acp{SDR} below it and the server to the left of it.}
    \label{fig:system_setup}
\end{figure}

The transmitter, receiver, and signal processing equipment are shown in \cref{fig:system_setup}.
All components are \ac{COTS} devices.

The server, that was used for signal processing and the data transfer from and to the \acp{SDR} is equipped with an Intel Xeon W-3245 \ac{CPU} and an RTX~2080~Ti consumer \ac{GPU}.

The \ac{SDR} equipment is composed of four Ettus \acs{USRP} N310 \acp{SDR}.
Furthermore, we use a single \ac{NI} OctoClock for a \ac{GPS} disciplined clock and timing source for our \acp{SDR}.
With this setup, all four \acp{SDR}, each equipped with four ports on the \ac{Tx} and \ac{Rx} path, behave like a single \ac{SDR} system with  sixteen synchronized physical antenna ports for both the \ac{Rx} and \ac{Tx} paths.

The 16 physical \acs{Rx} antennas are arranged as a \ac{UCA} with a radius of \SI{6.5}{cm}.
Uniform spacing is ensured by a ring retainer around the antennas.
The antennas operate in the \SI{2.4}{\giga\hertz} band with an omnidirectional radiation pattern and vertical polarisation.

For the transmitting users a single NAMC \ac{SDR} module\cite{NAMC-SDR} is used, which is also synchronized to the OctoClock. Attached to this \ac{SDR} module are six antennas, each of which represents a single transmitting user.
Each user antenna is installed on a tripod, allowing for conveniently adjusting location and height, relative to the receiving antenna array.
According to the data sheet, the antenna gain is \SI[qualifier-mode=text]{5}{\deci\bel\isotropic} at \SI{2.5}{\giga\hertz} and the antenna polarization is vertical.

\subsection{Transmission signal}

For data transmission we use \ac{SCM}. Similar to the \ac{NOMA} setup in \cite{9148869} each of the six users is equipped with a single transmit antenna, and all users are perfectly synchronized because we use a single \ac{SDR} module for all users.

The pseudo-random signal of each user is first \ac{QAM} modulated and then up-sampled by a factor of 16 and pulse shaped by a \ac{RRC} filter to reduce the signal bandwidth to a total of \SI{1.92}{\mega\hertz} and fit the system sampling rate of \SI{30.72}{\mega\hertz}.

To allow for cross-correlation-based synchronization on the receiver side, a Frank-Zadoff-Chu sequence is added to the signal of each user.

At the receiver, we use the same \ac{RRC} filter parameters in combination with the corresponding downsampling factor of 16 to recover the modulated signals.

In each transmission a total of 685 complex training symbols and 3840 complex data symbols are transmitted per user.

\section{Results} \label{sec:results}

In this section, we first demonstrate the performance of our proposed algorithm for different network dimensions. Then we evaluate the effectiveness of the measures proposed in \cref{sec:background} and finally we investigate the performance in the presence of noise.

In the following, each data point is an average of 50 transmissions, each consisting of 685 complex training symbols and 3840 complex data symbols modulated with \ac{QPSK}.
We only used the receive signal of 4 of the 16 antennas in our system. The reason is, that in a system where the number of users is smaller than the number of antennas, we can compute an optimal solution using a linear system, because the system is over-determined. Since the maximum number of users is limited by our available lab equipment, we instead limit the number of receiver antennas to 4 to assess a under-determined system.

The 6 users are consecutively numbered and each user has \SI{3}{dB} less transmit power than the previous user. For example, the fourth user has \SI{9}{dB} less transmit power than the strongest (i.e., first) user. By distributing the transmit power in this manner we can run simulations for users with a very low \ac{SINR}.

In addition to the (uncoded) \ac{BER}, we also give numbers for \ac{CBER} which was measured after applying a \ac{LDPC} code with a block length of 7680 and a code rate of $0.7$ (implemented in the Sionna library\cite{hoydis2022sionna}). The demapping was performed using the \ac{APP} algorithm, which used the ground-truth \ac{SNR} of the decoded user symbols as input.

The timings given in this section do not account for the copying of the data to and from the \ac{GPU}. In a practical system the previous processing steps would also be performed on the \ac{GPU}, so that the data would already reside on the \ac{GPU} and no copying would be necessary.

By using manual parameter sweeps, we managed to achieve good training performance by employing the Adam optimizer\cite{kingma2014adam} with a learning rate of 0.005, a training duration of 50 epochs and a batch size of 128. The relatively large batch size is influenced by the fact that the \ac{GPU} cannot be fully loaded for smaller batch sizes, which would decrease the throughput. We tried to optimize the \ac{MSE} loss.

\begin{figure}[tb]
    \centering
    \includegraphics[width=\linewidth]{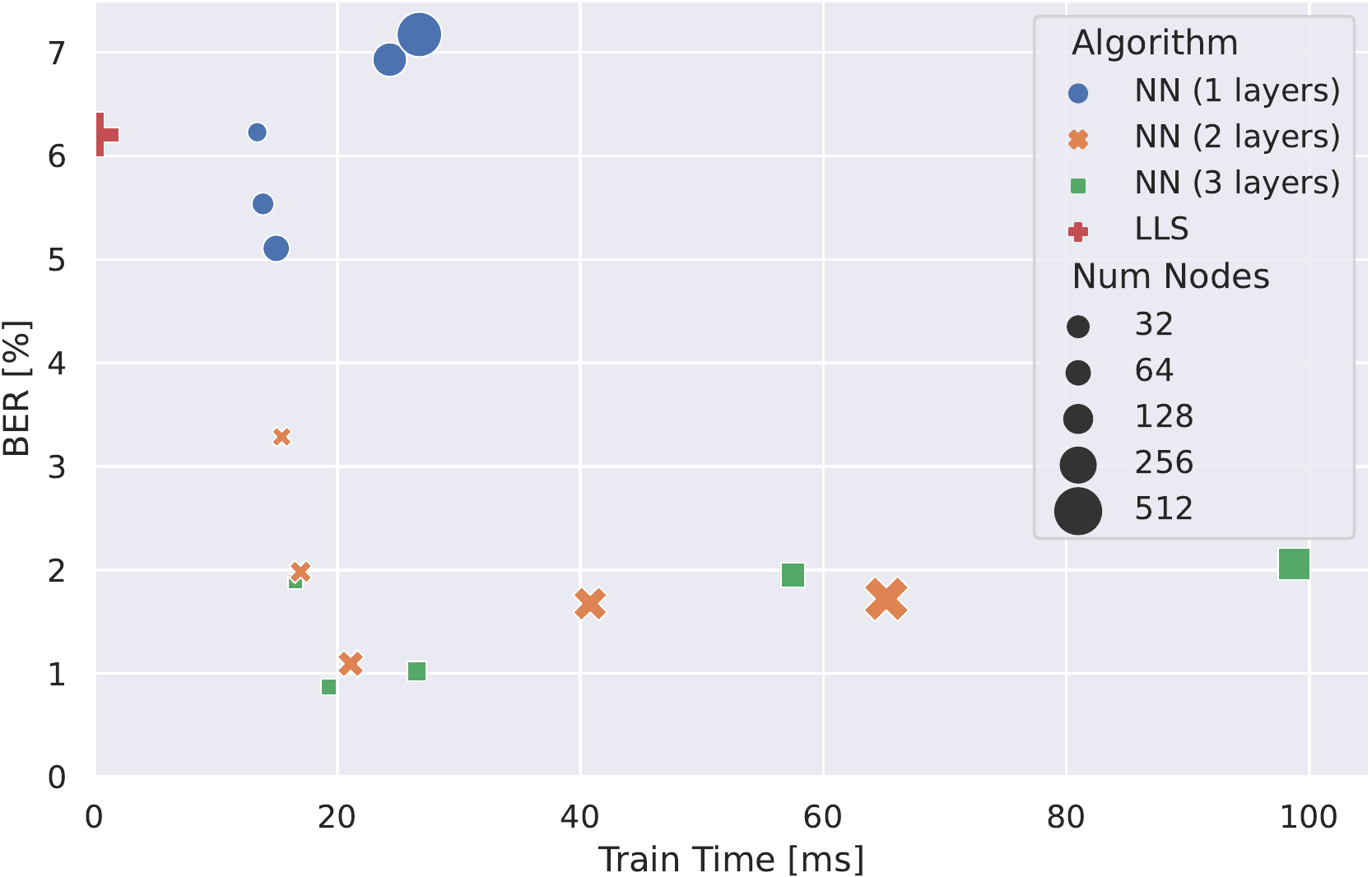}
    \vspace{-7mm}
    \caption{Average \ac{BER} over training time for different network dimensions.}
    \label{fig:ber_train_time}
\end{figure}

When choosing the optimal network dimensions, both detection performance in terms of achievable \ac{BER} and execution time need to be taken into account. The simulation results for different network dimensions are given in \cref{fig:ber_train_time}. A network with 3 hidden layers with 64 nodes each seems to provide a very good trade-off, as this achieves the lowest \ac{BER} and still has a very short training time. Numerical values for training and detection times are given in \cref{tab:execution_time}. We note that unlike for \ac{SIC}, where users are decoded sequentially\cite{verdu1998multiuser}, all users can be decoded in parallel and the execution time for our algorithm does not depend on the user that is to be decoded. The execution times given for Tensorflow were measured in graph mode on the same \ac{GPU}, but are not directly comparable, as they include additional overhead for Python wrappers and data copying.

\begin{table}[tb]
\centering
\caption{Average execution times of different algorithms.}
\label{tab:execution_time}
\begin{tabular}{lcc}
\toprule
     Algorithm & Training Time & Detection Time \\\midrule
     \ac{LLS} & \SI{238.7}{\micro\second} & \SI{13.4}{\micro\second} \\
     Fully-fused \ac{NN} & \SI{20.4}{ms} & \SI{32.7}{\micro\second} \\
     Tensorflow \ac{NN} & \SI{1.06}{s} & \SI{209.1}{ms} \\
     \bottomrule
\end{tabular}
\end{table}

\begin{figure}[tb]
    \centering
    \includegraphics[width=\linewidth]{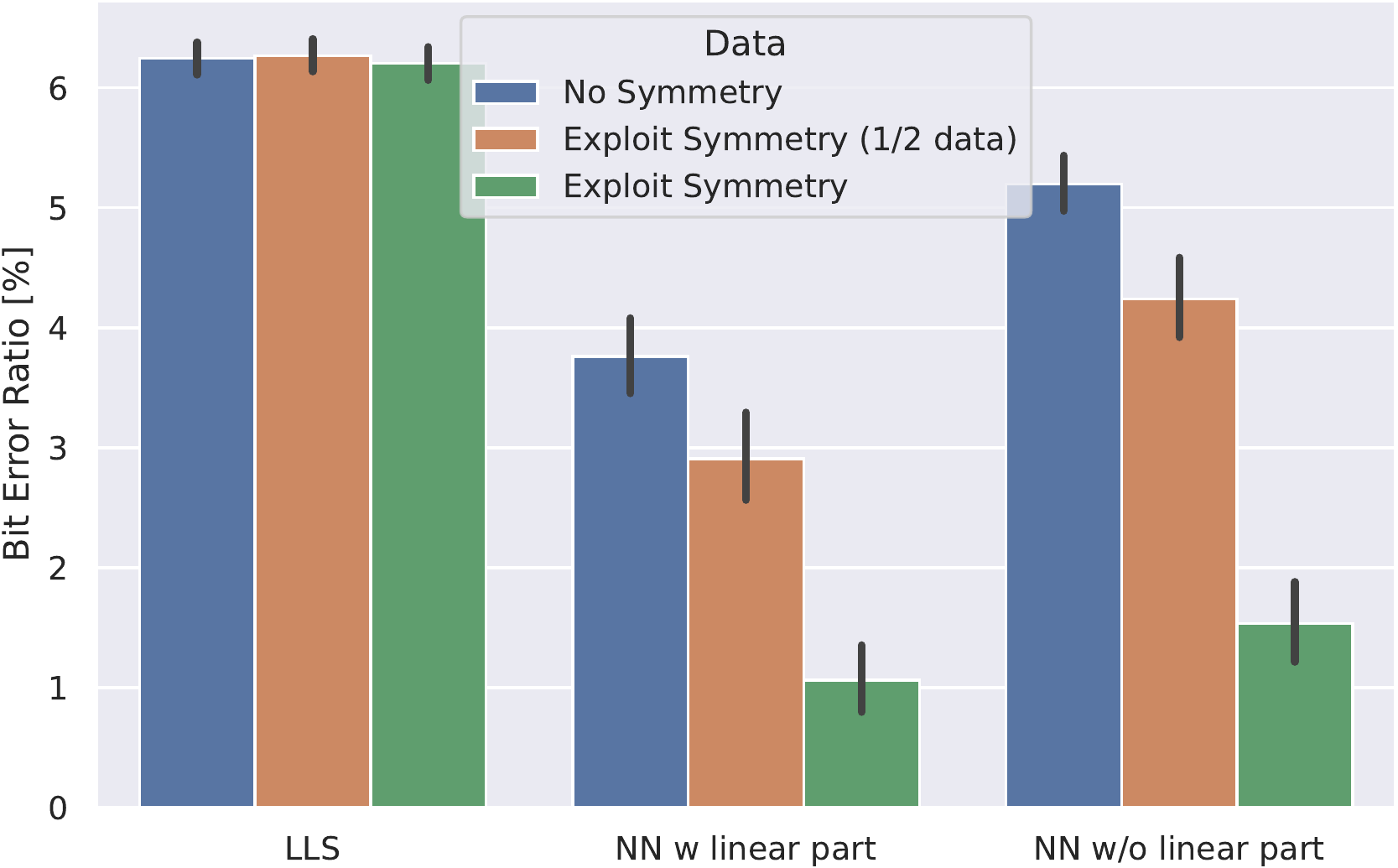}
    \vspace{-7mm}
    \caption{Average \ac{BER} of decoded user 4 when (not) exploiting complex symmetry. Error bars show the $\pm 1\mathrm{SD}$ confidence interval.}
    \label{fig:complex_symmetry}
\end{figure}

In \cref{fig:complex_symmetry}, we evaluate the performance gain that can be achieved by exploiting the symmetry of the IQ samples as proposed in \cref{sec:complex_symmetry}. For comparison, we also show the performance of the model, when being trained with only half the data set, so that the number of training samples is identical to training without this scheme. For the \ac{LLS} algorithm, the performance gain is negligible as the number of training samples is sufficient to achieve close to optimal performance for all schemes. For the \ac{NN} algorithm however, there is a big performance gain, which clearly demonstrates the benefits of this approach. The fact that the simulations with half the data set also show significantly improved performance compared to the original simulation, clearly demonstrates that the equivalence \eqref{eq:complex_symmetry} also holds for non-linear functions similar to our \ac{NN} and that the imposed constraints improve the performance of the network. Additionally, the data shows that adding the linear part significantly improves the performance compared to a conventional \ac{NN} structure with only a series of dense layers and activations.

\begin{figure}[tb]
    \centering
    \includegraphics[width=\linewidth]{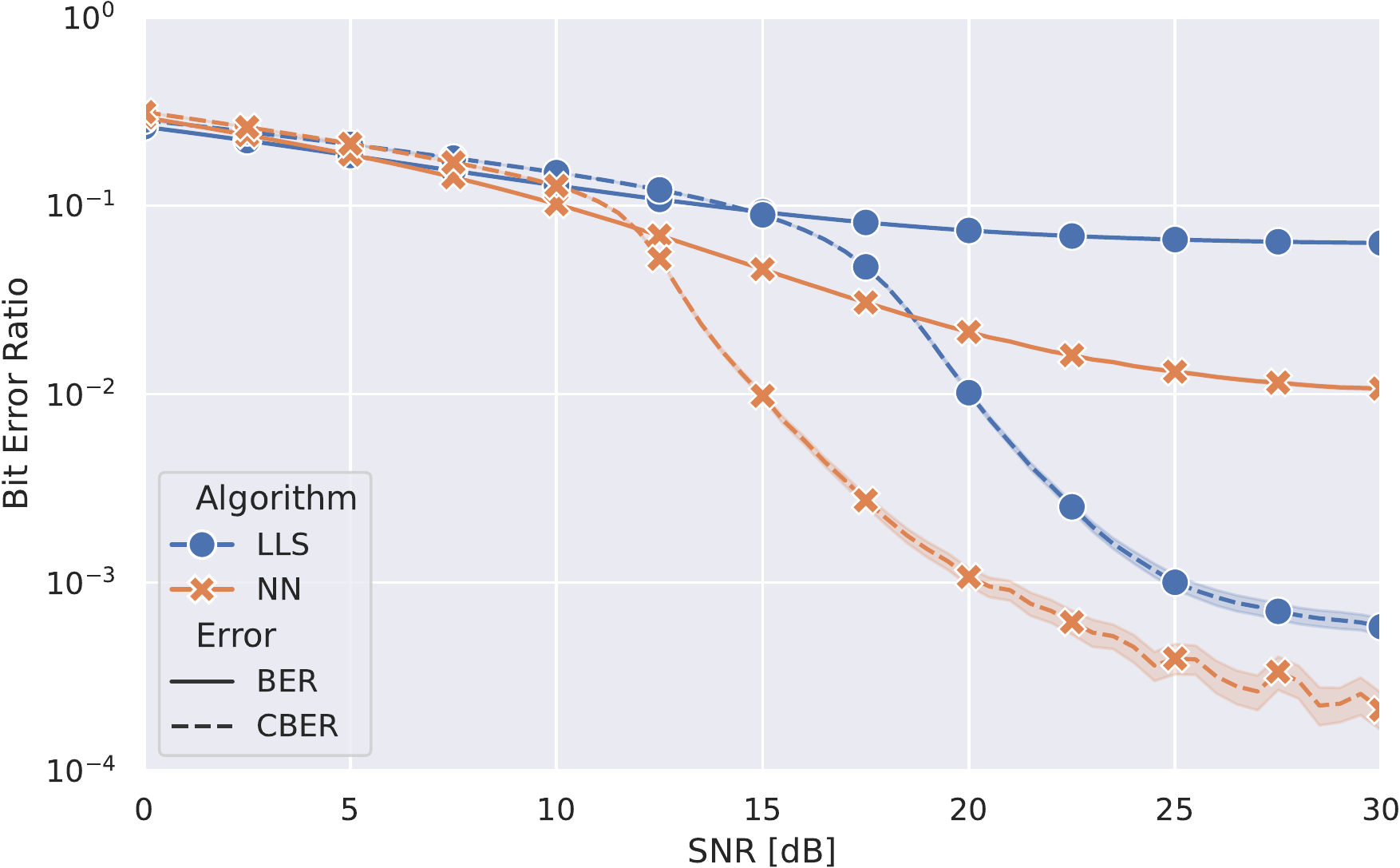}
    \vspace{-7mm}
    \caption{Average \ac{BER} and \ac{CBER} of decoded user 4 for different noise levels. Shaded regions show the $\pm 1\mathrm{SD}$ confidence interval.}
    \label{fig:ber_noise}
\end{figure}

In \cref{fig:ber_noise}, we examine the performance for different noise levels. To mimic the different noise levels we add artificial \ac{AWGN} to our recorded samples. The \ac{SNR} is defined here as the ratio of the power of the received signal (for all users and on all antennas) compared the power of the added noise. For each data point the simulation was run with 400 different noise realisations, giving a total of 20\,000 simulation runs per data point. The \ac{SNR} of our physical system was measured to be \SI{40}{dB}, which can be ignored for this experiment, as the artificially added noise is significantly stronger than the noise acquired by our measurement system. We note that the power of the \ac{SOI} (i.e., the received power of the fourth user, which we examine here) has a power of \SI{-12}{dB} relative to the received power of all users, so the \acp{SNR} for our \ac{SOI} are actually \SI{12}{dB} lower.

For \acp{SNR} above \SI{5}{dB} the \ac{NN} outperforms the \ac{LLS} algorithm.
The \ac{CBER} of the \ac{NN} drops towards an error floor about \SI{5}{dB} earlier than \ac{LLS}.
In the high-\ac{SNR} regime the \ac{NN} algorithm achieves almost an order of magnitude less uncoded \ac{BER}, and also a significantly lower \ac{CBER}. 
The error floor of both algorithms is most likely due to residual interference, which can occur due to non-linearities in the radio receiver or algorithms that are not powerful enough to remove all interference (possibly due to an insufficient number of training samples).

\subsection{Open Challenges}

We empirically observed that in the low-SNR regime, the \ac{NN} tends to overfit for some training cases. As a result, the learned \ac{NN} can not properly remove the multi-user interference. This can have an impact on the demapper, leading to a very asymmetric distribution of \acp{LLR}, which in turn will severely degrade the performance of the \ac{LDPC} decoder. In these rare cases, the proposed \ac{NN} can have a significantly higher \ac{CBER}, while still having a lower \ac{BER} than that of the \ac{LLS} algorithm.
We would like to emphasize that this challenge is mainly driven by the fact that the training must be done with a relatively small number of samples. However, it remains an open challenge to either find a more robust \ac{NN} architecture or training techniques to reduce the impact of overfitting while maintaining the average performance.

\section{Conclusion} \label{sec:conclusion}
We proposed to use \acp{NN} for \ac{NOMA}.
By adding a linear part with \ac{LLS} initialization and by exploiting the symmetry in the IQ samples, our proposed \ac{NN} structure allows for multi-user detection with very small error despite its relatively small dimensions. By utilizing an optimized, fully-fused \ac{GPU} implementation we achieve execution times that are several orders of magnitude smaller than those of a Tensorflow implementation and also small enough to perform the detection in the \SI{1}{ms} time-frame of typical \ac{ULL} systems. Having shown the feasibility of \ac{ULL} detection, in future research, we will address the training time, e.g., by tracking the channel and thus amortizing the training time across frames.

\bibliographystyle{IEEEtran}
\bibliography{literature}
\end{document}